\documentclass[sigconf]{acmart}
\AtBeginDocument{%
  \providecommand\BibTeX{{%
    \normalfont B\kern-0.5em{\scshape i\kern-0.25em b}\kern-0.8em\TeX}}}

\copyrightyear{2024}
\acmYear{2024}
\setcopyright{rightsretained}
\acmConference[CHI '24]{Proceedings of the CHI Conference on Human Factors in Computing Systems}{May 11--16, 2024}{Honolulu, HI, USA}
\acmBooktitle{Proceedings of the CHI Conference on Human Factors in Computing Systems (CHI '24), May 11--16, 2024, Honolulu, HI, USA}
\acmDOI{10.1145/3613904.3642402}
\acmISBN{979-8-4007-0330-0/24/05}

\begin{document}

\title{Data Ethics Emergency Drill: A Toolbox for Discussing Responsible AI for Industry Teams}

\author{Vanessa Aisyahsari Hanschke}
\email{vanessa.hanschke@bristol.ac.uk}
\orcid{0000-0003-3688-0372}
\affiliation{%
  \institution{University of Bristol}
  \city{Bristol}
  \country{United Kingdom}
}

\author{Dylan Rees}
\email{Dylan.Rees@lv.co.uk}
\orcid{0009-0004-2009-0805}
\affiliation{%
  \institution{LV= General Insurance}
  \city{London}
  \country{United Kingdom}
}

\author{Merve Alanyali}
\email{Merve.AlanyaliRafferty@lv.co.uk}
\orcid{0000-0002-2875-2656}
\affiliation{%
  \institution{LV= General Insurance}
  \city{London}
  \country{United Kingdom}
}

\author{David Hopkinson}
\email{David.Hopkinson@lv.co.uk}
\orcid{0009-0009-5163-4867}
\affiliation{%
  \institution{LV= General Insurance}
  \city{London}
  \country{United Kingdom}
}

\author{Paul Marshall}
\email{p.marshall@bristol.ac.uk}
\orcid{0000-0003-2950-8310}
\affiliation{%
  \institution{University of Bristol}
  \city{Bristol}
  \country{United Kingdom}
}

\begin{abstract}

Researchers urge technology practitioners such as data scientists to consider the impacts and ethical implications of algorithmic decisions. However, unlike programming, statistics, and data management, discussion of ethical implications is rarely included in standard data science training. To begin to address this gap, we designed and tested a toolbox called the data ethics emergency drill (DEED) to help data science teams discuss and reflect on the ethical implications of their work. The DEED is a roleplay of a fictional ethical emergency scenario that is contextually situated in the team’s specific workplace and applications. This paper outlines the DEED toolbox and describes three studies carried out with two different data science teams that iteratively shaped its design. Our findings show that practitioners can apply lessons learnt from the roleplay to real-life situations, and how the DEED opened up conversations around ethics and values.
\end{abstract}




\begin{teaserfigure}
  \includegraphics[width=\textwidth]{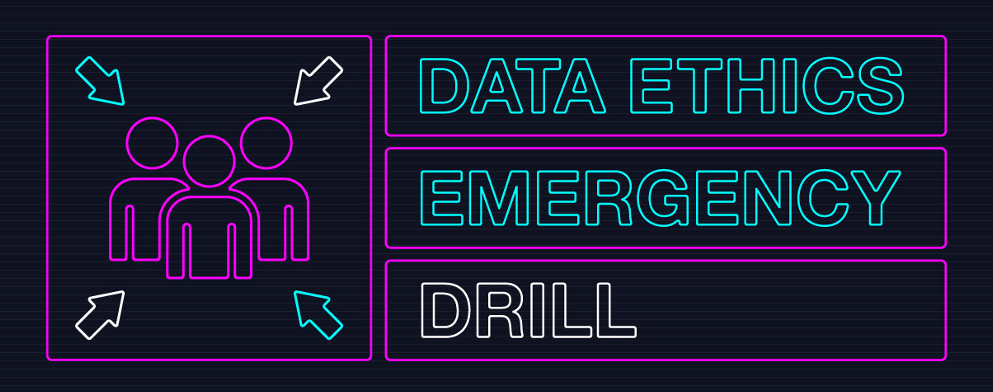}
  \caption{The Data Ethics Emergency Drill logo based on the icon for meeting point. The project website can be found at \href{https://www.gooddeed.ai}{www.gooddeed.ai}. }
  \label{fig:teaser}
  \Description{The meeting point symbol, which consists of three human icons and arrows pointing towards the group and the text of Data Ethics Emergency Drill to the right of the symbol.}
\end{teaserfigure}

\maketitle

%% ==========
%% INTRODUCTION
%% ==========
\section{Introduction}
Data science (DS), artificial intelligence (AI) and machine learning (ML) are rapidly expanding fields, and therefore teams of professionals specialising in these technologies have become commonplace in the last decade. As the importance of the data and AI industry grows, the work of these teams has an increasing impact on society. Recent studies\cite{boydDesigningValueSensitiveDesign2022,griffin2023ethical, passi2019problem} of data science, machine learning and artificial intelligence practitioners have shown that ethical decision-making is part of the everyday technical design choices practitioners have to make, but they are not always aware of it. In fact, discussing values outside of the technical remit of performance and considering societal needs has often been neglected in the field of AI \cite{birhaneForgottenMarginsAI2022}.

As these technologist teams have matured in their professional practice, some have incorporated AI ethics frameworks and techniques designed by researchers from fields such as HCI and STS into their processes to ensure that their products reflect their individual values and those of their company. However, when these research techniques are moved from the theoretical to the practical sphere, practitioners encounter the main limitations of current thinking about AI ethics: its abstractness and lack of context \cite{birhaneForgottenMarginsAI2022,wongSeeingToolkitHow2023}. This overlooks the crucial fact that data science, machine learning and AI in industry are
not themselves abstract concepts, even if this is how they might be discussed in research contexts. These technologies are being embedded into companies with various stakeholders inside and outside of the technical team and also into wider society, often replacing processes which have their own domain-specific norms that don't necessarily fit with general ethical guidelines.

 One proposed solution to bring more ethical sensitivity into the field is to educate technologists via discussions about ethics \cite{boydDesigningValueSensitiveDesign2022,griffin2023ethical}. Ethical roleplay has been extensively studied in the context of university education, including in computer science \cite{shapiroUsingRolePlayScale2021}. Examples show how roleplays can provide participants with a safe space to discuss and empathise with the decisions required of different stakeholders. Previous educational roleplays have confronted participants with generic problems of fictional stakeholders, and are typically presented within the context of formal education. However, there is an opportunity to examine real world scenarios with roleplays that are embedded into the specific context of a data science team. This means taking into account the complexities of the workplace, product lifecycle and societal impact.
 
 We designed a roleplay method called the Data Ethics Emergency Drill (DEED). These DEEDs ask questions such as: \textit{what if an algorithm that is in production is demonstrated to show bias against a group of people? What if malicious actors use your services? Would you know what to do and who to contact to resolve the issue, limit damage and prevent repetitions?}  The roleplays are intended to function like a fire emergency drill in the context of AI ethics and are designed in close collaboration with members of the data science team. The DEED method puts teams of data scientists into ethically difficult situations to explore and discuss the ethical implications of their work. By creating opportunities to confront an ethical issue within their day-to-day work, we aimed to encourage them to test processes and best practices which are in place, and to make values and ethical decisions explicit.

We conducted three iterations of the drill with two separate industry data science teams in different companies to understand two things: whether we could design data ethics emergency drills that are realistic and relevant to their work context, and how these would impact the practices of a team and the individual practitioners.

In this paper, we describe our iterative approach to designing the drills with members of the DS teams, the evolution of the DEED through these drills, and how we crafted contextual, embedded scenarios in the shape of online meetings. We describe the material that resulted from these studies and provide instructions on how to create a DEED for other DS teams to apply. Then, we present the initial feedback we received from surveying study participants and the insights DS teams were able to gain from their participation. We also present the results from follow up interviews carried out 9-15 months after the drill to investigate the long-term impact of the drill session. Finally, we discuss how we see the DEED being implemented, the limitations of our work and possible future evolutions.

Our main contribution to HCI research on responsible data and AI design is a new practice which incorporates consideration of and embedding into specific contexts. We demonstrate how this practice can impact the work of Data and AI practitioners, and we offer the DEED as a method that is part of this practice for practitioners to use, including a toolbox in the form of diagrams and instructions on how to use these.

%% ==========
%% BACKGROUND
%% ==========
\section{Related Work}

\subsection{Current Responsible AI and Its Shortcomings}
As AI, data science and machine learning technologies are being implemented across multiple industries, scholars have pointed out the dangers they can pose to society, such as exacerbating discrimination against minoritised groups \cite{crawfordThereBlindSpot2016, buolamwiniGenderShadesIntersectional2018, bender2021dangers}. In response to some of the questions raised by responsible AI researchers, solutions have been offered to support ethical AI governance, including ethical frameworks \cite{hagendorffAIEthicsIts2022, birhaneAlgorithmicInjusticeRelational2021}, AI regulation \cite{europeanparliamentandcounciloftheeuropeanunionRegulationEU20162016}, audits \cite{costanza2022audits} and checklists for practitioners \cite{madaioCoDesigningChecklistsUnderstand2020}. Technical solutions for explainability \cite{lundberg2017unified,ribeiro2016should} and debiasing methods \cite{zhaoLearningGenderNeutralWord2018} have been proposed to assist with transparency and fairness. There have also been proposals of design methods, for example using value sensitive design \cite{friedmanValueSensitiveDesign2006} to support the creation of responsible AI systems during the design phase {\cite{gebruDatasheetsDatasets2020, benderDataStatementsNatural2018, starkApologosLightweightDesign2021}}. Other design methods also support reflection and discussion in relation to ethical values with existing technological applications \cite{ballardJudgmentCallGame2019a,elsayed-aliResponsibleInclusiveCards2023}.

When implemented in real-life, some of these responsible AI solutions exhibit their shortcomings. The development of AI regulation has been criticized as being reactive and not suitable for the contextual complexities of algorithmic implementations \cite{hamonImpossibleExplanationsExplainable2021}. When asked to apply technical methods for explainability, data scientists have been found to put too much trust in interpretability tools \cite{kaurInterpretingInterpretabilityUnderstanding2020}. Furthermore, ethical guidelines may provide a fig leaf to cover up issues beneath the surface \cite{hagendorffEthicsAIEthics2020} or even be used to ``justify the path'' \cite{taylorConstructingCommercialData2020} set out by business models. In short, instead of foreseeing and preempting the ethical concerns pointed out by scholars, current solutions may gloss over responsible AI issues by either being too abstract and general or creating a false sense of trust. This focus on the abstract can be seen all across the field of AI ethics as Birhane et al. found in a recent survey of papers from FAccT and AIES \cite{birhaneForgottenMarginsAI2022}. Such revelations are not new: research in technology accountability has advocated time and again for supporting technology developers in considering their local impact and histories \cite{suchmanLocatedAccountabilitiesTechnology2002,katellSituatedInterventionsAlgorithmic2020,klumbyteCriticalToolsMachine2022}. This contextual awareness becomes especially important for AI applications, as they are distributed across different industries and interact with broader society. Besides the wider context of the application industry such as health care, social media, finance, etc., there is also a company specific context that shapes the responsible innovation practice of a data science team, including organisational structures, values and ways of working. In their analysis of 27 AI Ethics toolkits, Wong et al. \cite{wongSeeingToolkitHow2023} find that they lack instructions on how to implement ethical work inside organisational structures, which creates a "mismatch between the imagined work of ethics and the support the toolkits provide for doing that work" [p.1].

A number of studies addressed the question of what practitioners need from responsible AI by asking them directly. ML practitioners call for context-specific tools \cite{holsteinImprovingFairnessMachine2019} or ways of making contextual information evident in existing tools \cite{ashktorabFairnessEvaluationText2023}. They ask for tools that fit into their resource constraints, as ML practices outside of big tech may not have the bandwidth to carry out some of the responsible AI investigations of larger companies \cite{hopkinsMachineLearningPractices2021}. Finally, technologists require structures for discussing and reflecting on the ethical implications of their work \cite{boydDesigningValueSensitiveDesign2022,griffin2023ethical, holsteinImprovingFairnessMachine2019, doThatImportantHow2023}. Conflicts between values themselves and between different professionals will arise naturally, so to address this organisations will need to  ``(...) mobilize resources to create safe spaces and encourage explicit disagreements among practitioners positively [and] enable them to constantly question RAI values (...)'' \cite[p.13]{varanasiItCurrentlyHodgepodge2023}.

\subsection{Designing for Reflection with Ethical Roleplay}
We see ethical roleplay as an opportunity for practitioners to practice ethical discussions in a focused and efficient manner, and to lean in to the most challenging aspects of their specific applications.
Ethical roleplay has been extensively studied within the context of formal university education for areas including management \cite{agboolasogunroEfficacyRolePlaying2004,nandedkarEthicalDecisionMaking2019}, business \cite{brownUsingRolePlay1994}, accounting \cite{taplinUseShortRoleplays2018}, medicine \cite{nestelRoleplayMedicalStudents2007} and computer science and technology \cite{avinExploringAIFutures2020,louiWhatCanStudents2009,shapiroUsingRolePlayScale2021}. In most of these cases, students were presented with a scenario and were asked to play the role of different stakeholders. Avin et al. also applied roleplays for thinking about AI impacts in industry contexts, as they share in their ongoing research report \cite{avinExploringAIFutures2020}. They take inspiration from wargames which have long been applied in fields such as cybersecurity \cite{anderson1996exploration}.
The main motivations behind these studies were to create empathy in students for “implications for all stakeholders” \cite{nandedkarEthicalDecisionMaking2019}, prepare students for real-life situations \cite{nestelRoleplayMedicalStudents2007}, to introduce ethics into a largely technical curriculum \cite{shapiroUsingRolePlayScale2021} and to encourage self-reflection \cite{agboolasogunroEfficacyRolePlaying2004}.

We build on the concept of reflection in previous work of ethical roleplays. The importance of undergoing the process of acting out a scenario to acquire skills in handling ethical problems is highlighted by Taplin et al. \cite[p.385f.]{taplinUseShortRoleplays2018}: ``Ethics is best learnt by doing rather than being subjected to the traditional classroom transmissive teaching approach since the question of what is right or wrong is too abstract without the challenge of actual consequences.”
This active role of the learner can also be found in reflective design such as in Slovak et al. \cite{slovakReflectivePracticumFramework2017}, who make use of Schön's reflective practicum. The reflective practicum is “a setting designed specifically to generate particular sort of experiences that allow the students to explore by doing.” \cite{slovakReflectivePracticumFramework2017} Slovak et al. identify two main components: (1) teachable experiences as a base for (2) scaffolded processing to reflect and ultimately learn.
Our work draws on these principles of reflection, by using roleplay as a safe space for developing a deeper understanding of responsible AI and making abstract discussions concrete.

Previous work on ethical roleplay did encounter several challenges. Researchers struggled to create scenarios that felt personally applicable to the participants and students had a lack of knowledge about certain stakeholders (e.g. the role of a CTO) making it hard for them to suspend disbelief \cite{agboolasogunroEfficacyRolePlaying2004,avinExploringAIFutures2020}. If these ethical roleplays are not contextually relevant to their participants, then they fail to address the same gap as other solutions proposed by the AI ethics field: the method becomes too abstract. They also overlook the opportunity to engage participants in the concrete ethical dilemmas of their work.

To understand how we could craft contextually relevant scenarios, we were inspired by work from the Human Computer Interaction literature, participatory methods to create future narratives \cite{mullerExploringAIEthics2017} and methods which require participants to act out a role in a fictional scenario such as speculative enactments \cite{elsdenSpeculativeEnactments2017}, user enactments \cite{odomFieldworkFutureUser2012}, design fiction probes \cite{noortmanHawkEyeDeployingDesign2019} and Live Action Role Play \cite{pothongDeliberatingDataDrivenSocieties2021}. This research also involves participants in narratives and underlines the importance of enacting a situation, as opposed to discussing it in abstract terms. In particular, the concept of consequentiality, meaning that the enactments ``... generate circumstances where at least some elements or conditions of the speculation really matter to the participants'' \cite[p.5391]{elsdenSpeculativeEnactments2017} offers promising insight to the effect ethical roleplay could have in industry contexts of ML, AI and DS. The performance of acting out narratives in social contexts of mundane situations means participants have a stake in the scenario, which may address the abstractness that is often criticised in other futuring methods.

%% ========== '
%% Research Context
%% ==========
\section{Research Context}

\subsection{Author Roles}
The first author is the principal investigator of this project, which is her PhD project. Her  supervisor is the last author. The first author has an interdisciplinary academic background in linguistics, design and computer science and also has three years of industry experience working as a data and AI consultant. The second, third and fourth authors are data science practitioners in the financial sector. They have participated in the creative research process and offered feedback on the design and setup of the early drafts of the drill. The three practitioners were not involved in the data analysis of the findings. This was carried out by the PhD student and her supervisor. The last author is a HCI academic with a strong commitment to designing and evaluating technology in context, and experience of using narrative methods to explore ethical issues.

\subsection{Context of Participating Industry Teams}
The two participating teams were both data science teams based in the UK and recruited through University research showcases. One team was a large team of 50+ data practitioners set within a larger company. This data science team works on a range of projects implementing primarily supervised machine learning techniques into customer-facing and internal company processes. The other team was a small team with three data practitioners working within a startup sized company offering natural language processing services. The teams were recruited on the basis that they were interested in exploring their responsible AI practices. Before commencing we confirmed that their management was supportive of team members taking time away from work to participate in the scenario crafting workshop and the drill session.

\section{Method}

\subsection{What is a Data Ethics Emergency}
Our main idea was to confront data scientists with fictional, yet plausible dilemmas about the ethics, fairness or societal impact of their applications that needed to be solved urgently with the tools that they currently have at their disposal in their organisations. 
We call these data ethics emergencies. The literature shows there is no shortage of AI ethics frameworks \cite{hagendorffEthicsAIEthics2020}, not to mention ethical frameworks in general. Discussions involving ethics and values can happen at various levels (personal, product and organisational). Their resolution may or may not lie in the direct responsibility of the data science team, and can involve engaging with wider organisational structures \cite{varanasiItCurrentlyHodgepodge2023}. The ways organisations practise data ethics and their motivations not only vary, but are also entangled with one another \cite{taylorConstructingCommercialData2020}. Because we expected to encounter this variety and entanglement in our studies, we chose not to impose top-down structures or boundaries when brainstorming data ethics emergencies in our research approach. Instead we let the participants lead the discussion on what aspects of their responsible AI practice they were interested to explore based on their own team values. The analysis of the conversation topics in the findings mirrors this entanglement in the breadth of their content (see section \ref{sec:content_analysis}). Our aim was to find topics that had the potential to spark an engaging discussion about the embedded values of a system. To this end, we chose dilemmas which could be described as ``wicked problems" \cite{rittelDilemmasGeneralTheory1973}, i.e. they were not straightforward to solve and had trade-offs in their possible solutions. The data ethics emergency also had to be urgent, i.e. something that needed to be addressed immediately so that discussions would focus on concrete actions that could be taken. How these were generated is described in Section \ref{sec:idemergency}.

We identified  three types of desired outcomes for the DEED:
\begin{enumerate}
    \item To \textbf{test} if a data science team's \textbf{current processes} are adequate to address ethical emergencies, and to both understand what works well and what needs to be improved
    \item To offer a space for data scientists to \textbf{practise conversations around ethics} in their work context and to reflect on their practice both as a team and individually
    \item To \textbf{make values} embedded in the design of their technological applications \textbf{apparent} and reveal the impact of the design decisions on the wider society
\end{enumerate}

\subsection{Research through Design}

To understand if a situated role play could provide a good basis for discussion within technical teams and how to realise such scenarios, we opted for an iterative design process in close collaboration with practitioners, where we developed each subsequent study in response to participant feedback. 
Overall, we followed a research through design approach \cite{zimmermanResearchDesignMethod2007}, by trying to make the 'right' thing as opposed to the commercially viable thing. Our artefact in the design research sense is the DEED method, while the broader knowledge we would like to contribute to the responsible AI research community is a proposal for designing methods that allow data and AI practitioner teams to engage with the contextual details of their working environment.  

We had a lot of uncertainties around the drill design, such as how realistic prompts should be, the amount of detail they needed to contain, how extreme the stories in the prompts had to be to induce a sense of urgency and if the setting of a fake meeting would be natural enough to drive a discussion. It is through the process of designing these workshops with the practitioners that we defined our DEED method. The iterative approach allowed us to gain insight with a small number of studies and be respectful of our participants' time, as we were able to directly implement feedback. Practically, this involved, analyzing feedback that we gained from surveys and re-watching recordings of the drill, and using these insights to define the goals of subsequent studies. Finally, we identified important elements from both the scenario crafting workshop and the drills themselves so that we could create our toolbox of instructions for the DEED.

\begin{table}
    \centering
    \begin{tabular}{l | c | c | c}
         & Drill 1  & Drill 2 & Drill 3\\ \hline
         Number of scenario crafters  & 4 & 2 & 2 \\ \hline
         Number of drill participants & 6 & 6 & 4 \\ \hline
         Technical (t)  vs non-technical (nt) & 6t : 0nt & 5t : 1nt & 3t : 1nt \\ \hline
         Company & A & A & B \\
    \end{tabular}
    \caption{\textit{Number of scenario crafters} refers to the number of participants in scenario crafting workshop. This includes the researcher, but not company employees who were consulted for advice before and after the workshop. \textit{Number of drill participants} always includes one mole participant from the previous scenario crafting session.}
    \label{tab:my_label}
\end{table}

\begin{figure*}[ht]  
    \centering
  \includegraphics[width=\textwidth]{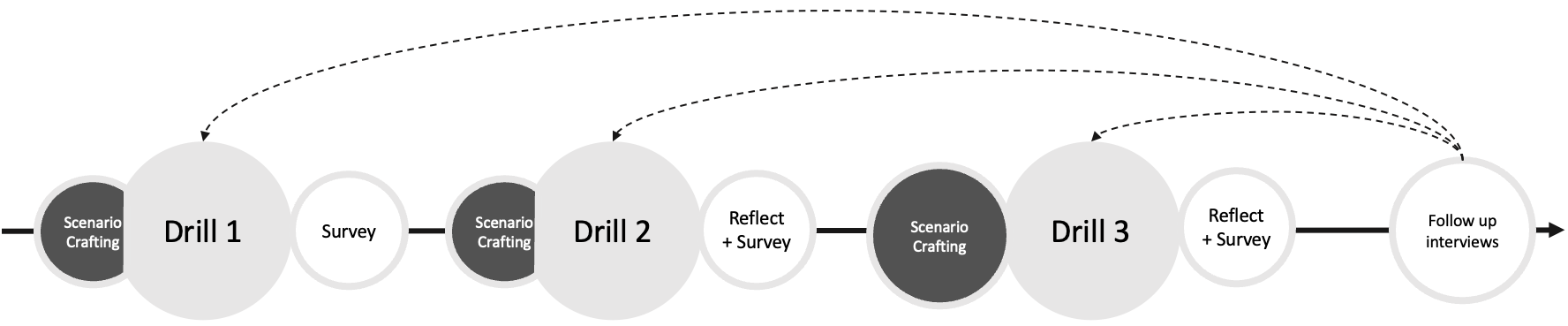}
  \caption{Overview of Drill Studies - The cut off circles for scenario crafting represent that these activities were incomplete before the third drill. The dashed arrows pointing from the follow up interviews to the drills represent that participants from all drill iterations were included in the follow up.}
  \label{fig:study-overview}
  \Description{A line with circles on it. The line represents the duration of the study and the circles depict different parts of the study. There are three clusters of studies for each drill each with a scenario crafting part, a drill part and a survey part. In the end, there is a circle for the follow up interviews with arrows pointing back at all the drills.}
\end{figure*}

\subsection{Designing the DEED with practitioners}

\subsubsection{Crafting the scenario}  We sought to design a tool that was resource efficient and educationally effective while being contextually relevant to the DS team's area of practice. Therefore, we involved members of the team in the crafting of the scenario, which included brainstorming dilemmas, structuring the narratives, and planning how to implement the drill in their teams.  
The 
scenario crafting workshop 
was held with one to three members of each DS team before executing the DEED. During these sessions the first author asked detailed questions about potential issues concerning responsible AI inside their practice, how these could vary, and how they would unfold in real life. The members of the DS team and the researcher then identified suitable candidates for scenarios and 
discussed who would need to be involved to conduct them. 
The researcher then created multiple drafts of scenarios and the final one was reviewed and chosen by a member of the DS team. 
The researcher then crafted prompt material for the drill session such as fictional news material, fake internal emails or fake presentations.

\subsubsection{Executing the drill session} We organised online meetings with a different group of members of the same DS team to participate in these scenarios. The researcher stepped out of the meeting, so as not to disturb or distract from the discussion and re-watched the drill from a recording. The scenario was structured by prompts that were designed in the previous workshop and participants would discuss their reactions as if in a real work meeting. The prompts would either be sent in the form of fictional emails from colleagues, that were sent to their work inbox, or colleagues joining the call to report a problem. Thus, all the resources used to create the prompts were not only simple to make, but also appropriate for the work environment of the DS team. Group dynamics and job roles were replicated in the drill scenario structure and participants "played" themselves. In doing so, participants were able to probe their understanding of their individual responsibilities in the case of such an emergency and who they would expect to interact with and ask for help.
After the drill, participants provided feedback in the form of a post-session questionnaire or in discussions after the session.

\subsection{Evaluation - Data Collection and Analysis}

To evaluate our DEED method, we analysed each study right after we conducted the drill and conducted follow-up interviews about a year after the drill to understand the long-term impact.

\subsubsection{Post-Session Evaluation}

Feedback was collected from all 16 participants in the form of online questionnaires, which were completed individually in the week after the drill session took place. 
The survey included closed multiple-choice questions, as well as free text open questions where participants could provide more detailed accounts of their experience. The main goals of the survey was to understand if the drill scenario and setup was a realistic depiction of their work context and how this affected the discussion.
We used the responses of the surveys along with the drill transcripts and the materials produced in the drill (e.g. notes and slide decks) for our analysis.
The first and last author carried out a thematic analysis \cite{braunUsingThematicAnalysis2006} of the open responses in the survey. The drill transcripts and materials were analysed using content analysis \cite{eloQualitativeContentAnalysis2008a}. The first drill was analysed inductively to establish the four main categories. The subsequent drills were then analysed deductively along those four categories. Finally,  we conducted a global analysis of the themes within the four main categories across participants of all three studies.

\subsubsection{Long-term Evaluation}

After about a year (9-15 months) of the studies taking place, we reached out to participants who had consented to being contacted again. Seven participants were available to be interviewed. Each of the three studies had at least two respondents participating in a follow-up interview. The participants had varying professional roles and had participated to different extents in the drill. Some had participated in the crafting of the scenario, some in the drill itself and some in both. To gain more honest feedback, the interview was carried out by a colleague from the same institution as the first author, but who had previously not been involved in the study. This was to discourage participants from shielding their answers as a result of not wanting to offend the chief investigator. The seven interviews were each between 35-60 minutes, held online on Microsoft Teams and recorded through Teams. The interviews were semi-structured covering how the participants remembered the drill, what the action points/learning were, if any of the action points had been implemented, if any similar discussions to the drill scenario have occurred and if they had seen any difference in ethical literacy\footnote{Ethical literacy was defined as the capability to discuss ethical issues and identify relevant aspects in participants' own projects.} in themselves or their teammates.  The interviewer also showed a document of action points that had been presented back to the whole data science team after the drill to understand what had been implemented.
For these interviews, we offered 20 GBP multi-retailer gift vouchers to remunerate participants for their time, as this was an add-on to the previous study held within their team and participants may have changed roles in the meantime.
The video recordings were transcribed and subsequently analysed using thematic analysis by the first and the last author. 

\subsection{Research Ethics}
The studies were approved by our university ethics committee. To ensure that participants were not affected in their work environment and relationships, we established a protocol which included how to exit the discussion or end the meeting if they felt uncomfortable by contacting the researcher through chat or email. We reminded participants at the start of the drill and the next day in a follow-up email that all the prompts were fictitious and that the DEED is designed to be a reflection activity to improve processes, not a test of their personal capabilities. Participants were given reassurance that the drill will not affect their performance evaluations or employment status as well as keeping their participation anonymous to the wider team.  No complaints regarding work relationships were recorded during the drill or in follow-up surveys.

\section{Iterations of different Drill Structures}
Below is a description of each drill that we conducted and the high level learnings that shaped each iteration. We have omitted commercially sensitive details of the drills and precise role descriptions to keep participants' identities anonymous.

\subsection{Drill \#1: Team A - Email Prompts}

The first drill was a one hour online meeting with six participants in the roles of data scientist or data engineer and one volunteer from the leadership team of the DS team. The volunteer from the leadership team had been involved in the design of the scenario and was assigned the role of the "mole" for this first pilot study. The "mole" was in touch with the researcher via a Microsoft Teams chat to ensure that prompts had been received and that the discussion had finished before the researcher re-entered the call. Participants received three prompts in the form of emails, with 20 minutes to discuss each prompt. The email prompts outlined a problem found in an existing model implementation with increasing severity. The prompts were sent by the researcher but signed by the model monitoring team, a manager and from a public facing department calling for action. 

This drill showed that we were able to craft an engaging discussion with the simple tool of fake emails.
Some feedback suggested that including a business stakeholder in the discussion or even letting them present a problem directly in the drill would make the scenario more realistic.
When analysing the transcript, we found that some of the learning goals we had set were not addressed in the discussion, and survey responses about the participants' takeaways were lacking detail. We thought this could be supported by explicitly asking participants about their understanding of the scenario and how the results of the drill could be turned into actions.
Another important insight from this drill was the importance of the "mole" who we initially put in place to coordinate prompts with the researcher. We observed in this study that the mole often served as a driver of the discussion and was filling in missing information in the prompts.

\subsection{Drill \#2 : Team A - Relay of Presentations}

The second drill was for same data science team as the first, but with a different group of six participants, including a new "mole", discussing a different problem. The most substantial change from the first study was that prompts were through presentations by the participants themselves rather than email, including a non-technical stakeholder from the business side. Moreover, we added a reflection activity in the form of a questionnaire to support reflection on the drill. The questions were based on our insight from the first drill. 

To replace email prompts with presentations, the second drill was designed as a relay of three separate online meetings. In the first, a business stakeholder presented a problem to managers. In the second, one of the managers told a group of data scientists and data engineers the problem. The group then had to prepare a presentation for the business stakeholder based on the plan made by the managers beforehand. In the last meeting, one of the participants from the second meeting presented the detailed plan to the business stakeholder. 

The involvement of the business stakeholder delivering the prompt as a presentation instead of email and the reflection activity all helped generate insightful discussions and were added to the DEED toolbox. The presentation that data scientists made to present their action plan turned out to elicit a very focused discussion and was a useful artefact for analysis, as the main discussion points were already summarised.

\subsection{Drill \#3 : Team B - Mixed Presentations}

The third drill was held with four members of a data science team at a different company. This iteration was held with the main purpose of testing the drill with a different team in a company of a different size. Some additions were made to the scenario crafting workshop: as the researchers were new to the context of this company, another activity was added to map values and practices of this company that had not been included in the first scenario crafting workshop.

This online meeting had two prompts delivered via email and one prompt delivered in person by a participant, since both in person and email delivery had been judged as realistic in previous iterations. The drill included the reflection activity, which had proved to be useful in the second drill.

%% AN EXAMPLE OF A DRILL
%% ==========

\section{Illustrative example}
\label{sec:IllExample}
The following fictional example illustrates what a drill scenario looks like. This toy example shares many characteristics with the actual drills that we designed, which cannot be put into this paper as they are commercially sensitive. These characteristics are:

\begin{enumerate}
    \item the problem discusses a 'wicked issue' where solutions require trade-offs i.e. how do you model a heterogeneous ever-changing group of people
    \item the problem addressed a responsible AI value in their team
    \item the problem appears to be caused by a realistic data ethics issue i.e. resulting from data or an algorithm using data
    \item prompts are in the form of written information from people within the company
    \item prompts increase in severity or difficulty over time
    \item prompts have clear directives on what to discuss
    \item prompts involve a variety of stakeholders
\end{enumerate}

 This example is inspired by recent research that has found bias against second language speakers of English \cite{markl2022language} in automatic speech recognition systems (ASR).

\fbox{\begin{minipage}{.9\linewidth}

\textit{The drill plays out in an online meeting where around five data science team members with various roles (data scientist, data engineer, manager, etc.) gather in an allocated time slot. In this example, the technical team is working on a service using ASR to engage with customer calls.  The participants receive the following three fictional emails over the course of an hour, which structure their discussion during this meeting:}

\begin{enumerate}
    \item\textit{\textbf{Prompt 1:} The scenario begins with an email from customer relations. They have found that there has been a rise in complaints from immigrant citizens using their service since the introduction of ASR. They ask the DS team to investigate.
    Following this email, participants discuss ways they would probe the training data, the model and the customer log to understand the issue.}
    
    \item\textit{\textbf{Prompt 2:} An email from the data science manager. She had a discussion with the data provider, and their data does not include representations of non-native speakers. She asks for a presentation of potential solutions for this problem.
    Participants discuss ways that the model could be improved or the training data could be extended. They discuss what might be short- and long-term fixes to cater to non-native speakers through customer calls.}

    \item \textit{\textbf{Prompt 3:} An email from the CEO who loves Twitter. A celebrated AI ethics researcher noticed the difficulty their migrant father has when using the service and posted about it on Twitter, relating it to general pitfalls of automation. The CEO asks for input to provide a technically sound public response.
    Participants discuss what they would include in the response, such as their team's reasoning behind the design decisions that have caused the problem and their intention to fix it as soon as possible.}
\end{enumerate}

\textit{After the discussion, the participants individually answer a set of written questions about their drill to prompt them to reflect on their experiences. What would have been their first action stepping out of the meeting? What, in their opinion, was the main issue in the drill? What would their response to the tweet in prompt 3 have looked like? What learnings do they take from the drill for their specific role in the company?}
\end{minipage}}

%% ==========
%% DEED
%% ==========
\section{Suggested Anatomy of a Data Ethics Emergency Drill}

In this section, we describe how to structure a successful DEED, including the preparation work of crafting the scenario \textbf{before the drill}, running the session itself \textbf{during the drill} and how to draw insights for the drill from the experience \textbf{after the drill session}. We provide reflections and considerations based on what worked in the three drills we ran. A complete set of workshop slides can be found in the supplementary materials. The figures and the materials were visually embellished after the studies in collaboration with a graphic designer.

\subsection{Choosing Participants}
It is useful to first establish the participants or the participant pool to know participant numbers and who can be involved in the preparation of the drill. Our past iterations have always involved at least one participant in the creation of the scenario, while the others were confronted with the scenario on the day of the drill. Participants were people within the data science team or colleagues from outside the team who work closely with the data scientists.
We recommend a group including a minimum of one participant from the DS team involved in crafting the scenario and a minimum of three participants for the drill. For the drill, the group discussion should be limited to around six people to allow for enough space for each participant to talk. If more people want to participate, then the scenario crafter could consider splitting these into separate sessions. We found it useful to designate one "mole" amongst the drill participants, who was aware of the goals of the discussion and who could communicate with the person coordinating the prompts in the background. They can also direct participants' attention to important parts of the prompt and give believable answers to questions related to the hypothetical scenarios. We found managers have a natural suitability for such a scenario, as they might usually drive discussions in real-life meetings.

\subsection{Before the Drill - Crafting the Scenario}
% Designing the Scenario
This section describes a summary of the steps to create the scenario and craft the prompts in a workshop with data scientists. 

\begin{figure*}[p]  
    \centering
  \includegraphics[width=\linewidth]{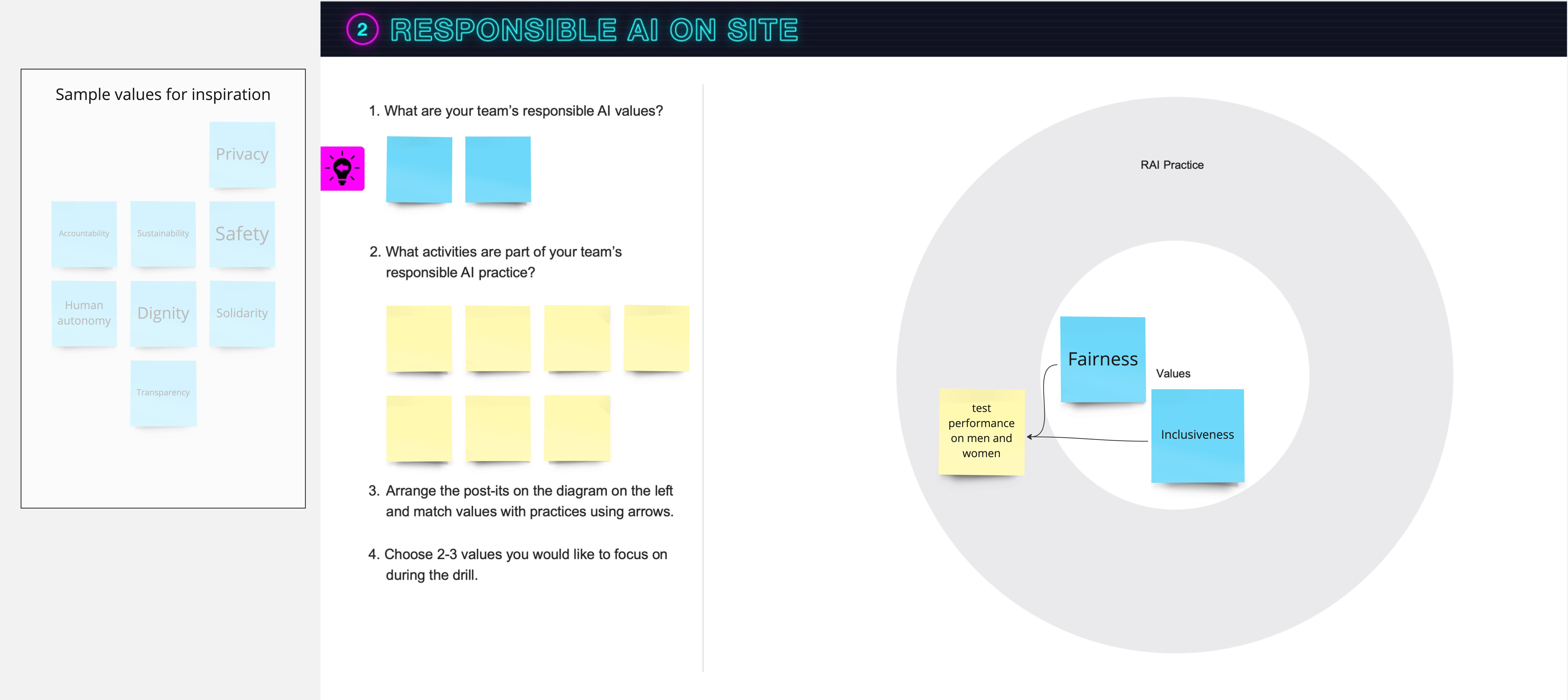}
  \caption{Screenshot of example of values activity on the Miro board}
  \label{fig:values}
  \Description{On the left a collection of post-its with the sample values: privacy, accountability, sustainability, safety, human autonomy, dignity, solidarity and transparency. In the centre there are instructions of the values mapping activity with empty postits. On the right two circles, one contained inside the other. The inner circle is labeled values the outer circle is labeled responsible AI. There are two post-its in the values circle labeled fairness and inclusiveness. They are connected with arrows pointing towards a posit-it with the text: test performance on men and women.}

\vspace*{\floatsep}
\hfill
\hfill
\hfill

  \includegraphics[width=\linewidth]{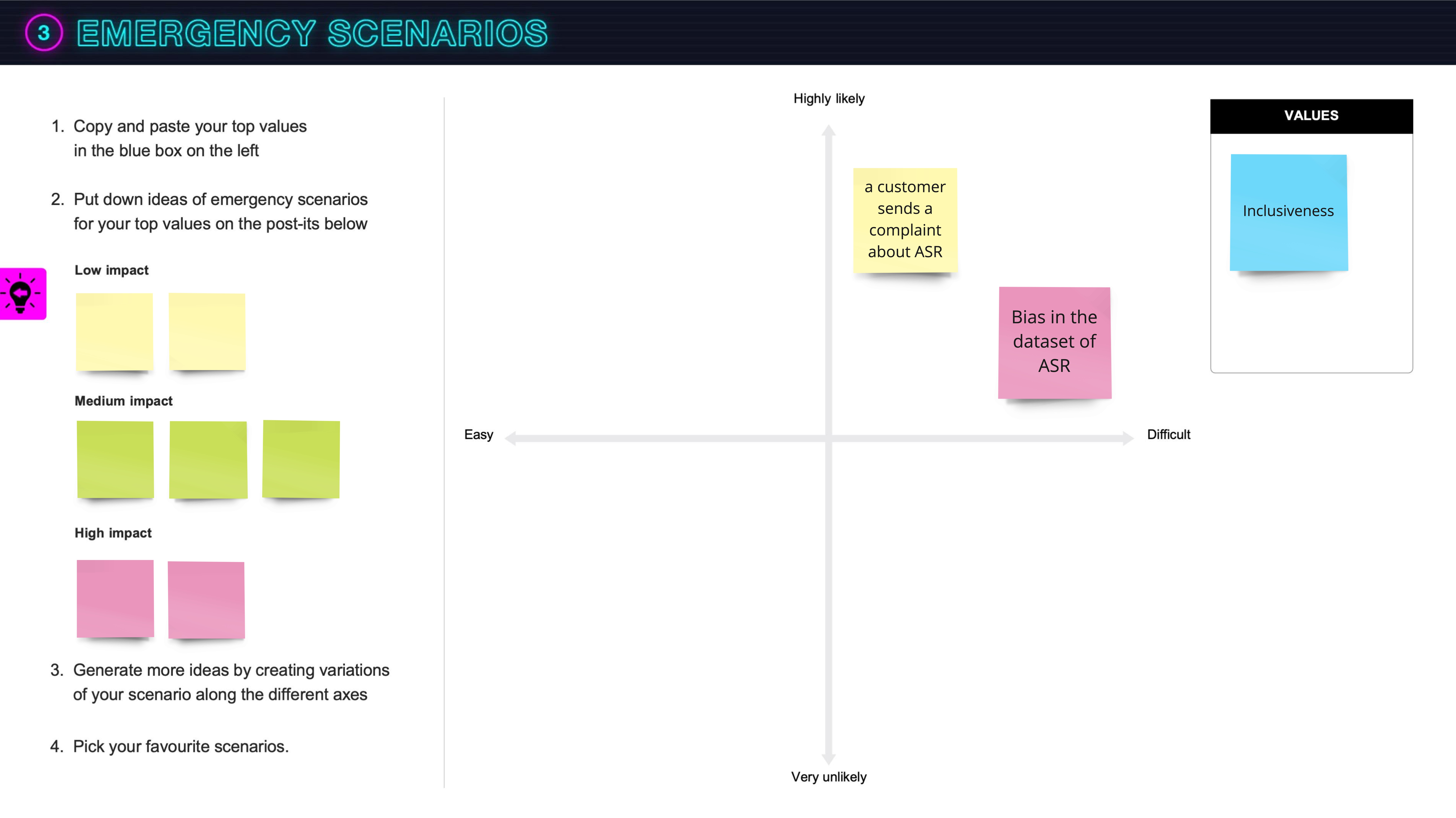}
  \caption{Screenshot of example activity to identify emergencies on the Miro board}
  \label{fig:scenarios}
  \Description{A graph with four quadrants and multi coloured post-its. The y-axis goes from very unlikely to highly likely. The y-axis of the graph is labeled from very unlikely(bottom) to highly likely(top). The x-axis is labeled easy(left) to difficult(right).  In the upper right quadrant there are two post-its with scenarios, one saying a customer sends a complaint about ASR and the other Bias in the dataset of ASR. To the right of the post-its are the instructions to the activity for brainstorming emergency scenarios. To the left, there is a blue box with the chosen value inclusiveness.}
\end{figure*}

\begin{figure*}[ht]
  \centering
  \includegraphics[width=\linewidth]{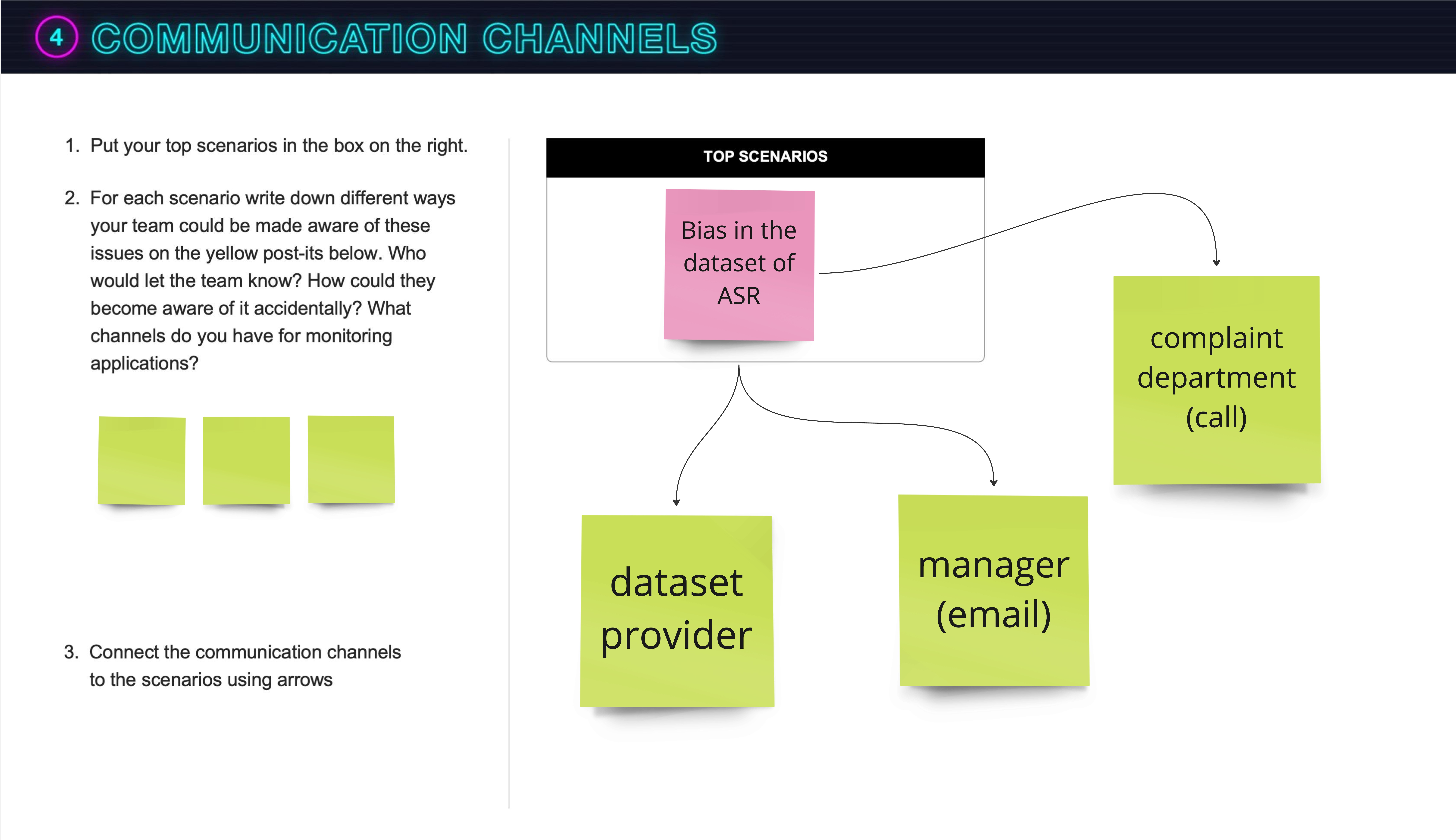}
  \caption{Screenshot of example of activity for identifying communication channels on the Miro board}
  \label{fig:communications}
  \Description{A box labeled top scenarios is on the top of the image with a post-it in it that says bias in the dataset of ASR. Three post-its are outside of the box saying dataset provider, manager (email) and complaint department (call). The instructions for the communication activity are on the right hand side of the picture.}
\end{figure*}

\begin{figure*}[ht]
  \centering
  \includegraphics[width=\linewidth]{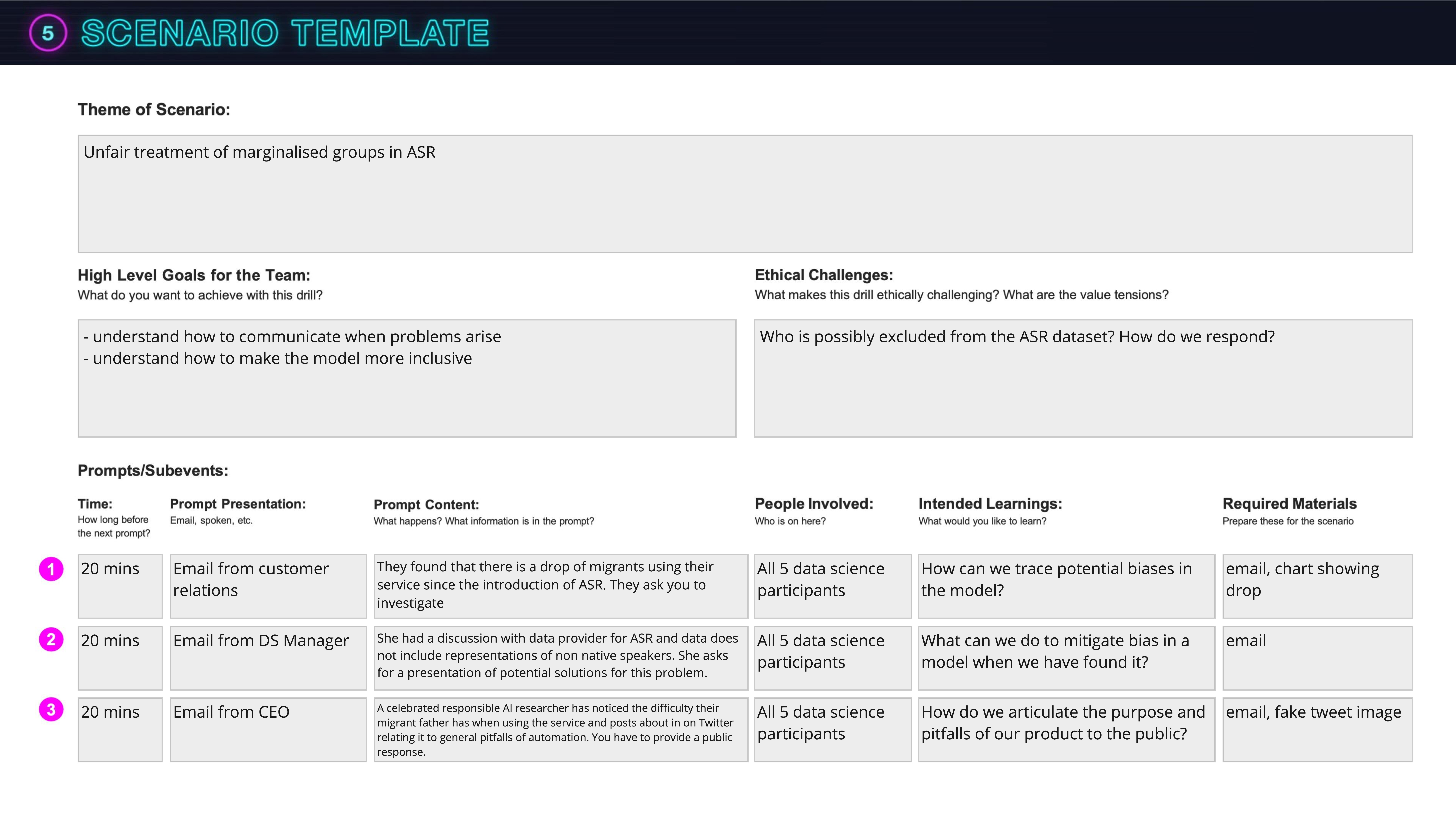}
  \caption{Screenshot of example activity for structuring the drill on the Miro board}
  \label{fig:structure}
  \Description{A form with several filled text boxes. The titles of the boxes are: theme of scenario, high level goals for DS Team, Ethical Challenges and Prompts/Subevents. The scenario structure is filled in with the example from the Illustrative Example section of the paper.}
\end{figure*}

\subsubsection{Mapping of Responsible AI Values and Practices}
\hfill\\
\textbf{Task:}
Participants describe responsible AI values for their team and the processes that are currently in place to support these values. Values are placed in the inner circle and processes in the outer circle. Then, the processes are connected to the corresponding values. We found it helpful to initially start discussion without prompts, but also have a set of values pre-written if the team does not already have an established list of values or struggles to list values . Figure \ref{fig:values} shows the Miro board used for this activity. This exercise is to establish the current state of responsible AI in the data science team and to get an overview of the maturity of the processes. We expect that this discussion can already inspire some of the potential data ethics emergencies. 

\textbf{Discuss:}
Which of these values are most important? Do any of the values have intrinsic tensions? Have any of the processes had problems?

\textbf{Examples:} Ethical principles from \cite{jobinArtificialIntelligenceGlobal}, e.g. transparency, fairness and sustainability.

\subsubsection{Identifying Emergencies} \label{sec:idemergency}
\hfill\\
\textbf{Task:}
Participants take 2-3 of the established values they would like to focus on and place them on the board shown in Figure \ref{fig:scenarios}. Then participants brainstorm potential ethical emergencies. These are placed along the axes of difficulty to resolve and likeliness. They are also marked using colour for three levels of impact (high, medium, low). Participants should try to cover all areas of the graph and think of versions of the same problem on different dimensions i.e. easy vs difficult, likely vs unlikely, low impact vs high impact. Generating versions of different magnitudes of difficulty, likeliness and impact can help create an overarching narrative when making the prompts into a coherent drill scenario. In the end, participants choose 2-3 problems that they would like to focus on. 

\textbf{Discuss:} What are common criticisms of the DS/ML/AI application/technologies? How might current events affect applications, data or the model? Who is excluded from the model or data? 

\textbf{Examples:} 
\begin{itemize}
    \item External events cause data drift in the data set.
    \item A model exhibits bias in a category that was previously not tested. 
    \item Public criticism of the systems use of data and AI
\end{itemize} 

\subsubsection{Communication Channels}
\hfill\\
\textbf{Task:}
Based on the chosen problems, participants map how internal or external communications will reach their team to make them aware of the chosen set of problems. This includes understanding the communicators, the means of communication and the time between the problem occurring and the news reaching the data science team. Figure \ref{fig:communications} shows the Miro board used for this activity. This exercise serves to directly structure the scenario and design the discussion prompts, by collecting information on how these could be appear in the drill scenario.

\textbf{Discuss:}
How would the team be made aware of such issues? Are the news sources outside or inside of the team? What channels does the team have for checking use to monitor applications? Who are the main informants and what is their relationship to the team?

\textbf{Examples:}
This could be colleagues responsible for monitoring the application and the models in place. It could also be colleagues directly in touch with end-customers such as the complaints department or communications. It could also just be a line manager. 

\subsubsection{Structuring the Drill Scenario}
\hfill\\
\textbf{Task:}
Participants lay out their narrative using the structure in Figure \ref{fig:structure}. This includes an overarching theme, the goals for this drill, what ethical challenges it addresses and what internal processes or practices are being tested. Prompts can be delivered indirectly, e.g. the informant sends an e-mail, or in person, i.e. the informant attends the meeting. There are practical trade-offs in deciding between in person and email delivery of prompts, while in person offers the possibility of participants interacting with the informant, this requires the informant to be available for the drill and to be comfortable "acting". We found that this acting role suited some informants in the team more than others and that emails also successfully prompted engaging discussions. 
The overall structure of the drill should see problems getting increasingly more difficult and urgent to solve, so that discussions can build on top of each other. For each prompt, the intended learnings define the overall theme of the scenario. Giving participants action points that are explicit in the prompts, will help guide the discussion (for example, in an email prompt the sender can ask participants for help with an aspect of the problem). The prompts should include a sufficient amount of detail to feel realistic, but also consider what excuses can be made for leaving out details (e.g. access to the data set, which technical teams love to scrutinize). There are online tools to make convincing prompts such as FakeTweetMaker.com \footnote{https://www.faketweetmaker.com/} or simply sending emails to oneself to create a convincing looking email chain. It is important to ask consent from real people included in fake prompts before including their email signatures. This exercise serves to clarify roles and goals of the crafted scenario.

\textbf{Discuss:}
 How much discussion time will each prompt need? Do the prompts form a coherent narrative? How should people presenting prompts be involved? Does each prompt present enough new information to stimulate a discussion?

\textbf{Examples:}
See Section \ref{sec:IllExample}.

\subsection{During the Drill - Executing the Drill}
% Drill Execution
To start, the participants are briefed about the general setting of the drill and the code of conduct. Our general setting was a work meeting which is interrupted by outside events, which then take precedence over the discussion. The code of conduct should include the company code of conduct and generally establish the educational nature of the drill (as opposed to evaluating the team). It should ask participants to suspend disbelief during the session, but emphasise that all the events and prompts are fictional.
Then the drill starts and the prompts are presented in succession. In our studies, this was coordinated by the researcher, who did not participate in the drills. The researcher/drill leader keeps time and sends out any prompts that are designed to be emails in the background. They can coordinate with one assigned participant, dubbed "the mole", who gives feedback if prompts are received and if it is an appropriate time in the discussion to send the next prompt.
Participants can choose to take notes or produce artefacts such as email responses, presentations, etc. to record the main topics and insights of their discussion. This can be decided based on whatever comes naturally to participants or is appropriate for the length of the drill. We highly recommend taking notes or producing some artefact, as these not only force participants to be concrete with their ideas, but also help focus the discussion. Furthermore, written artefacts serve as an aid for analysis after the drill. During our studies, the sessions were always recorded for the researchers to analyse, but notes produced in the sessions seemed to cover most of the points that the researchers found through their analysis of the recordings.

\subsection{After the Drill - Reflecting}
% Reflection Activity
\label{sec:AfterDrill}

After the drill, participants answer a reflection questionnaire. The reflection questionnaire is to support participants in reflecting on their drill experience and expressing their learnings. It captures:
\begin{itemize}
    \item any questions that the drill was supposed to answer, which may or may not have come up in the discussion (see drill structure)
    \item how the participant saw their role
    \item how the participant interpreted the purpose of the drill
    \item what changes the participant would like to see in their team based on the drill
    \item what questions were left open
\end{itemize}
This questionnaire also contributes to scaffolding reflection \cite{slovakReflectivePracticumFramework2017}, as it allows participants to revisit their experience and link it to concrete process change and actions that can be undertaken in their role at the company.
A sample can be found in the supplementary material.

Finally, the team needs to summarise and communicate the findings of the drill. The reflection activity offers a good basis for analysis and gathering insights. Notes and artefacts such as presentations are other rich materials for extracting insights and we also recommend re-watching the drill to capture anything that is not represented in the other material. While some of the action points might be very obvious from the reflection activity, we found it useful to list all the points and to categorise discussion topics into four types: 
\begin{itemize}
    \item Concrete Action Points: things that would be immediate actions when the emergency arises. These may often only be applicable to the emergency itself, but could indicate if some steps need to be added to more general procedures. Example: Check data source has no errors.
    \item Future Action Points: this would not be a direct response to the drill, but rather something that the team would implement once the emergency has been dealt with. These are often the most useful, as they represent a process change or a preventative measure. Example: Create guidelines for the team when bias is found in a model that is in production.
    \item Open Questions: questions that were unanswered during the drill. Often they are wicked problems and require further discussion. Example: What is an acceptable threshold for bias in this application?
    \item Clarification: questions to ask people outside of the drill. These are potentially interesting for creating more realistic crafted scenarios.
    Example: Ask customer relations about the exact volume of complaints they have received.
\end{itemize}

Based on these categories, the DS team can then decide what necessary actions to take and what needs to be discussed further.

%% ==========
%% Results from three studies
%% ==========

\section{Findings}

In the following section, we present the findings in three parts: the participant feedback directly after the session, the content analysis of the drill recordings and the evaluation through long-term interviews.

\subsection{Participant Post-Session Feedback}

\begin{figure}[ht]

  \centering
  \includegraphics[width=0.8\linewidth]{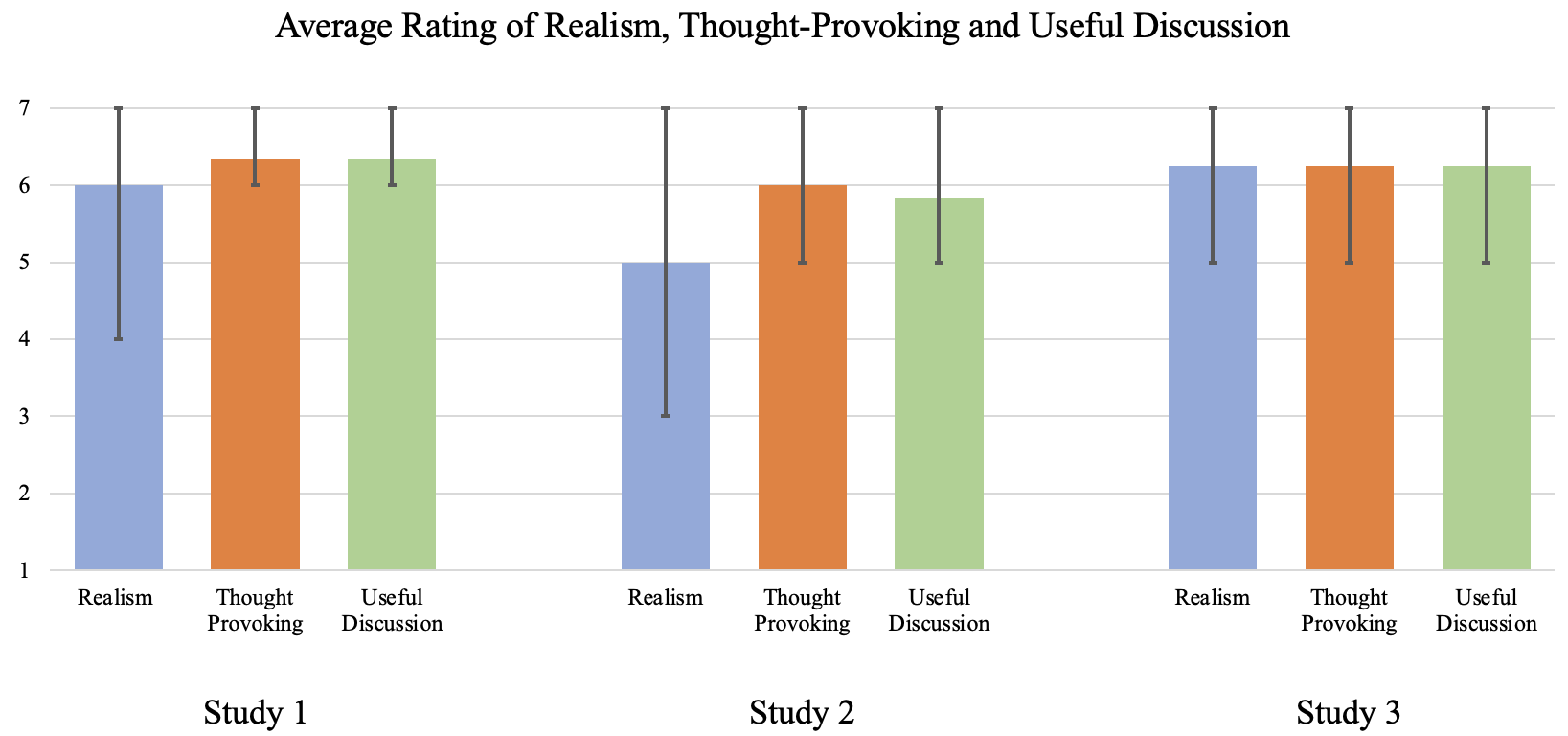}
  \caption{Average Ratings of each study for perceived realism of the scenario, ability to provoke new thoughts and to elicit a useful discussion. The bars show the mean ratings and the black lines mark ranges of lowest rating and highest rating.}
  \label{fig:ratings}
  \Description{Bar chart of mean ratings of realism, thought provoking and useful discussion for the three studies also showing the ranges of each category. Most categories have a score of around 6 or higher, except for realism in Study 2 which is at 5 with responses ranging from 3 to 7.}
\end{figure}

Participants scored the studies highly across all categories with a mean of 5.75 for realism,  6.2 for being thought provoking and 6.1 for eliciting a useful discussion. (Answers were given on a 7 point scale with 1 being the lowest and 7 the highest score.) The detailed mean scores between the three studies are shown in figure \ref{fig:ratings}.

Our main goal of the initial feedback was to understand how the realistic context shaped the discussion and participant experience and how to improve the set up. Participants found the setup of the drill as a meeting with email and presentation prompts largely convincing, natural and realistic for their work context. When asked about the importance of \textbf{realism} for the drill, most participants said it \textbf{was helpful for immersion}, e.g. \textit{``The realism framed the problem as a work problem to solve instead of just a theoretical problem''}.
An area of contention between participants around realism was that those in technical roles would have liked to have had access to fictional data sets. These data sets were referenced in the drill, but did not exist in reality. Also, they would expect more involvement from stakeholders who weren't present in the drill. Addressing both of these criticisms comes with practical trade-offs. Depending on the scenario, it might not be straightforward to create the fictional data set and depending on the team structure, it might be logistically complex to involve a large number of stakeholders. We also discussed during our scenario crafting workshops that we run the risk of distracting participants from the actual goals of the scenario, i.e. discussing the fictional ethical dilemmas, if they start putting effort into analysing a fictional data set. The positive feedback shows that these missing details of the scenario setting did not inhibit provoking new thoughts or usefulness.

An unforeseen positive outcome that came through in survey responses was how the drill shaped the relationship between participants. Many participants from the three studies described a \textbf{positive effect on confidence in the team after the drill}. ``\textit{...[E]veryone was able to contribute. I have faith that my team members would act diligently if this sort of scenario were actually to take place.}''

\subsection{Content Analysis of Discussion Topics of the Drill}
\label{sec:content_analysis}
As described in \ref{sec:AfterDrill}, we first focused on categorising the discussion points in a way that the data science team could make use of them: concrete action points, future action points, open questions and clarifications. We presented the analyses of the points to the participating teams. After collecting these points from three studies, we conducted a global analysis of the themes within the four categories across participants. Numbers in brackets indicate the number of discussion points that were recorded across the three studies that fit into the descriptions.

\subsubsection{Concrete Action points} A large number of points (10) addressed understanding the problem and its origin better, where a subset (4) also involved understanding if the responsibility was within the data science team (i.e. if their technical solution caused the problem) or if it was extrinsic (e.g. the underlying data changed or old processes unknowingly had the same issues). All drills had points (4) discussing technical fixes for the problem presented in the drill. Several (5) included involving external people in a solution, such as the customers, suppliers or collaborating researchers.

\subsubsection{Future Action points} Several suggestions (4) for future action were made to add further monitoring mechanisms that would ensure certain issues wouldn't arise. Some (4) future action points involved updating formal processes such as pre-release checklists or creating new formal processes for specific issues. A couple of future action points discussed spending more time to understand if similar risks could apply at the start of projects. Lastly five points called for more awareness and knowledge of potential issues, for example through training, establishing guidelines for developers, or leadership advocacy around data ethics.

\subsubsection{Open Questions} Many of the open questions (12) were around accurately quantifying categories or concepts which are not straightforward to define, situated and socially constructed, such as fairness. Two points in separate studies came up, where participants asked what a proportionate amount of effort would be to put into such hard questions, which ultimately seem unsolvable. Some open questions (4) addressed company identity and some (4) professional identity, i.e. what tasks should be within the remit and responsibility of technical staff, as opposed to other members of the company. There were a few (3) open questions around ways of making recourse easier for end-users potentially affected by their applications.

\subsubsection{Clarifications} These mainly concerned the scale of the problem (2), the information source of the problem (3) (i.e. how the problem was picked up) and more details about the hypothetically affected customers of the drill scenario (4).
\hfill\\

We can see that the drill elicited discussions which covered a the breadth of topics and touched on a variety of levels of solutions starting from technical investigation of the data to organisational values to addressing bigger questions such as quantifying fairness. In the drills, as in real life, these aspects of ethical problem solving are intertwined. The drill seems to bring this entanglement to the fore in the relevant organisational context of the participating teams.

%% ==========
%% Long term impact of the drill
%% ==========
\subsection{Long-term Evaluation}

In the following section, we present the findings gained from follow-up interviews carried out with participants 9-15 months after the drill session. Quotes have been redacted for clarity. P1-P7 denote randomly assigned participant numbers.

% ---------
% Theme 1
%---------

\subsubsection{The DEED provided a safe space for exploring realistic dilemmas}

One team experienced a similar situation to their drill scenario, shortly after conducting the drill. The scenario had at one point questioned the role of developer responsibility when implementing customer requests that clashed with their values. Even though the details of the value clash and the part of the system in question were not the same in real life, they were able to directly draw on their ethical discussion skills  from the drill.

\begin{quote}
\textit{
``I think because we've sort of done the drill, (...) we've already got quite well versed in having that discussion and sort of knowing what to do. (...) How to handle that as an organisation instead of feeling like it was on one person's shoulders to make that decision.}'' - P3
\end{quote}

This quote also shows that, in this case, the drill shaped the understanding of shared responsibility in the organisation. Because of their previous hypothetical discussion, the participant recognised that this was an issue worth escalating. Comparing the real-life situation to the drill, the participant also explained how each experience was different. P3 continued to suggest that the drill had been successful at creating a safe space to practise conflict, while still being relatable and transferable to real-life issues.
\begin{quote}
\textit{
``...I think we definitely learned a lot from [the drill]... you know, more so, than if it obviously happens in the real world you learn slightly different things but I think there's a natural, almost a paralysis around decision-making in the real world scenario because you can't afford to get anything wrong. It was nice to have a bit of safe space to experiment with how you might deal with things.}'' - P3
\end{quote}

Three other participants described how the projects they are working on have similarities to the drill. They discussed how they have approached their data with a deepened understanding of social and environmental factors that could impact their systems or be impacted by their systems, which is why they put additional monitors in place to account for these. For example: 

\begin{quote}
\textit{
``But rather than have the problem come to us, (...) we're kind of hopefully beating it by having monitors in place and sound reasoning behind our default...}'' - P2
\end{quote} 

These examples show how the DEED has influenced the responsible AI practice of participants in the real world.

% ---------
% Theme 2
%---------

\subsubsection{The DEED opens up discussions and sharpens awareness to unarticulated responsibilities}

Several participants (5) reported taking more initiative in opening conversations about ethical issues and taking the discussion outside of the data science team.

\begin{quote}
\textit{
``The thread of that conversation is carried on in the company and it spread to the [Higher Executive] and like in general what we think about our ethics sort of position around what we do.''} - P7
\end{quote}

\begin{quote}
\textit{
``I know better how to kind of lead the conversations and what things to pick up on.''} - P5
\end{quote}

\begin{quote}
\textit{
``And I think understanding that you know it's our (...) responsibility to raise those issues when we see them and push for them to be thought about because the likelihood is that people probably won't think about it unless you do something. Like at the end of the day, (...) we are like the experts in the data. (...) [W]e spend more time with thinking about this stuff than anyone else in the company probably.''} - P4
\end{quote}

These quotes show a deepening of participants' understanding of how their specific roles can contribute to those discussions, even though ethical discussions may not be seen as part of the day-to-day job of technical roles. At several points, they portrayed the typical data science way of working as number and solution focused, for example when saying: \textit{``yet another data science problem to solve''} (P1), \textit{``numbers on a screen and (...) model outputs''} (P2), 
\textit{``engineers(...) not particularly used to handling conflict resolution''} (P3) or \textit{``"as technical professionals, everyone was looking for (...) an answer''} (P6). The drill was seen to counteract that by giving \textit{``a bit of a shake''} (P1) to participants, for example by switching the lens to outside of the team:

\begin{quote}
\textit{
``It clicked a bit more where they're like: Oh well if this is gone like onto social media (...) And I guess as technical professionals, I know we're not necessarily subject to that kind of external lens.''} - P6
\end{quote} 

In this sense, the drill made participants articulate values in their data science systems that may not explicitly have been talked through before as they were not part of usual technical discussions. This gave way to a reordering of priorities as displayed in the following quote:

\begin{quote}
\textit{
``I think as (...) a team, we're sort of mostly scattered amongst like lots of different things all of the time (...). (...) [The drill] probably ended up, you know, shifting the priority of things and making sure things got done a bit more quicker maybe and helped by kind of goodwill in the rest of the team or (...) people with buying into getting certain projects over the line that were already underway''} - P4
\end{quote} 

This description of a shift in priority was echoed by other participants as a shift in \textit{``focus''} (P2) or making known issues \textit{``even more obvious''} (P3). One participant also describes this as a linking of their previous ethical education to technology.

\begin{quote}
\textit{
``I understand like ethical concepts as a whole, but then the application of that to like data tech engineering is something that I maybe pursued more after the DEED.''} - P6
\end{quote}

Another participant talks about widening their understanding from theoretical harms to how they would address these practically in their organisation.

\begin{quote}
\textit{
``Prior to doing it, [I had] an understanding of what things we could do, [who] we could end up discriminating against but not an understanding of what we would do if we find it, who to kind of speak to about it, what what we could do to prevent it. So I think that's where the DEED came in. ''} - P5
\end{quote} 
We see that by confronting practitioners with concrete, urgent data ethics dilemmas, participants were able to link abstract values and principles to the practical manifestations in their organisations. Discussing how issues would be addressed in a group gave them a clearer idea of how to navigate their organisational structures.

\subsubsection{The DEED confronts people with a diversity of opinions, but it could do more}
\label{sec:themeconfront}

Some participants reflected on the experience of being confronted with differing opinions in their team. 
\begin{quote}
\textit{
``You think you have kind of a view of the world or like a set of beliefs or like an opinion on a problem. But it's not until you actually sit down and discuss that and get people pushing back on it, and like people to bounce ideas off, that you really like flesh out where your position lies. ''} - P4
\end{quote} 

\begin{quote}
\textit{
``We came away like I understood what [Another Participant Name]'s position was. I understood what [Another Participant Name]'s position was, and we'd explored a lot of ideas. So we definitely came away with better vocabulary around it and an understanding of how the team feels.''} - P7
\end{quote}

\begin{quote}
\textit{
``I've kind of walked around with the perspective that I thought everyone was on the same page as me for a while (...)... there was some little things that maybe identify to me that not everyone was where I was at and even I wasn't where I wanted to be.''} - P6
\end{quote}

The three quotes show how the DEED offers the space to reflect deeply on one's own standpoint in relationship to other team members. When asked about repeating the drill within their team, one participant also saw the DEED as an opportunity to reinforce and share their own values, as seen in the next quote:

\begin{quote}
\textit{
``If we kind of have the responsibility that you have to really avoid bias and not have anything really along those lines. (...) 
%[F]or the most part, I think people in our team would would agree on that. 
So I think just on that basis from my own kind of ethical standpoint, I would want people to be aware of it and knowing what to do.''} - P5
\end{quote}

Some participants pointed out that they would have liked more input from outside of the technical team, to also challenge the common beliefs of technical people and bring in their knowledge.

\begin{quote}
\textit{
``I felt that as an engineering team, typically we have similar views anyway like we're quite similar individuals. (...) So it might be good to get those opposing views a little bit earlier on and then (...), it encourages little bit more discussion, a little bit more knowledgeable discussion as well.''} - P3
\end{quote}

\begin{quote}
\textit{
``[Bringing in a diverse group of people] can identify things that [business] stakeholders as domain experts, may be aware [of], so they can, you know, bring like a different point of view to the discussion rather than, like I  said, us data scientists just because it's so easy to just lose track (...) cause you just like shift to your comfort zone, which is the technical discussion''} - P1
\end{quote}

\subsubsection{The DEED should be part of an ongoing discussion}
While all participants had some positive takeaways, whether it was personal or organisational change, some participants (3) reported that they would have liked a more continuous discussion, to be able to enact some change.

\begin{quote}
\textit{
``...it was quite like finite in terms of like you could turn your brain on to it and turn your brain off of it.''} - P6
\end{quote}

Although the researcher had presented the findings of the drill afterwards and the list was circulated among the team, not all participants seemed to be aware or involved in this discussion. One suggestion that several (4) people brought up was having a longer series of meetings with, for example, a feedback session to the wider team and a group reflection on the findings of the drill.

 \begin{quote}
\textit{
``I might schedule in a meeting at the end to discuss with anyone who might be relevant and lots of people to keep it free and just to have it come in together later on to to say, this is the problem.''} - P3
\end{quote}

\section{Discussion and Future Work}

\subsection{Towards a Context-Specific Understanding of Responsible AI Practice}

We have not seen a silver bullet for responsible DS, AI and ML, and we believe that is not and should not be the end goal of responsible AI. 
As our findings show, solutions to ethical issues are deeply intertwined with organisational structures, product context and broader questions about value tensions.
Ethical literacy is a process, not an end goal and products should continually be monitored for value alignment. Similarly to Georgieva et al., we see the future of responsible AI as a ``landscape of methods, standards and procedure'' \cite{georgievaAIEthicsPrinciples2022}.
Our work demonstrates the value of incorporating context into responsible AI practice, and is a novel addition to this landscape. In our method, we sought to give DS teams several points where they could reflect on how ethical dilemmas apply to their specific contexts: when designing the drill scenario, when reacting to the prompts during the drill and when writing up their reflections afterwards. Not only did they step out of the technology-oriented routine of their delivery-focused mindset for the DEED, but they also showed how they were able to insert those elements back into real work conversations. This link between theory and real, contextual, organisational life is what has been highlighted as a major deficiency of current fairness toolkits \cite{wongSeeingToolkitHow2023}. Focusing on situated agency is in line with research considering response-ability \cite{klumbyteCriticalToolsMachine2022} and located accountabilities \cite{suchmanLocatedAccountabilitiesTechnology2002}.
We suggest that the key elements to the success of our method are three-fold: introducing friction, setting our method in a consequential environment and offering a safe space to explore responsibility beyond role definitions.

\subsubsection{Friction}
Unlike checklists and frameworks, the DEED very much encourages DS teams to take action and confronts them with areas of friction and value tensions, as opposed to offering passive acknowledgement of principles and paying lip service to AI ethics. This idea is similar to the use of ``values seams'' in the Apologos design method \cite{starkApologosLightweightDesign2021} to make implicit human values of technology explicit. By framing ethical dilemmas as realistic drills, the DEED tries to make evident which developments are necessary and urgent and centres the focus on the agency of the members of the data science team and how their services directly affect people. We can see these effects in the responses to the long-term interviews, where participants show how the drill made them reflect on their role and make ethics a priority in discussions.  

\subsubsection{Consequentiality of the Drill Setting}
By replicating relationships and roles in the drill session, the drill becomes consequential in a similar way to speculative enactments \cite{elsdenSpeculativeEnactments2017}. The setting of the drill is \textit{real}, in the sense that participants expose their own values and reasoning, which will shape relationships and conversations in the future. Hence, conversations during the drill have \textit{consequences}. For example, P7 expressed this in \ref{sec:themeconfront} when they said that they understood the other participants' positions and P6 when they described how this shaped their self-understanding. This understanding was then brought into future, real-life conversations.

\subsubsection{Mapping Responsibilities to Roles within the Organisation}
Modularity of software processes has been found to be a factor that may disconnect AI developers from accountability for their system \cite{widderDislocatedAccountabilitiesAI2023a}. A successful drill exposes the single modules and their interfaces, helping participants develop a better understanding of the responsible AI processes of the team as a whole. It may be that hierarchies and roles dictate, who will be exposed to certain ethical dilemmas in the day-to-day (e.g a manager will be more used to making high-level decisions than a junior data scientist). By connecting these roles in a scenario and creating a space where different roles and levels contribute to a discussion, we create transparency for different team members of what could or should happen when addressing ethical dilemmas. The contextual detail and realism of the scenario is crucial here in generating a useful discussion that practitioners can integrate into their work process. Furthermore, the fictional quality of the drill helps alleviate what P3 called the ``paralysis around decision making in the real world scenario because you can't afford to get anything wrong'' and creates a safe space for junior members to express their opinion.

 \subsection{Inserting the DEED into other Practitioner Teams}

After carrying out three studies, mainly motivated by our research goals and the curiosity of data science teams, we can reflect on how we see this work organically being implemented in other DS/AI/ML teams. The first opportunity we see is using the DEED as a team-building activity, because of how it boosted confidence in the team and served as practice for addressing ethical issues. If the team is large enough, then subgroups could design DEED scenarios for each other to encourage people to not only engage with different projects across the team, but also with their most challenging aspects. Secondly, the DEED has the potential to be a powerful tool in making a case for ethical design. This work could help any stakeholder in the pipeline demonstrate the issues they see with an Al application and underline the urgency of process change.

AI ethicists, team leads or values advocates, as studied in value-sensitive design \cite{shiltonBlendedNotBossy2017}, could take the lead in designing the scenarios and guiding the discussions. The DEED could readily be integrated with other frameworks, checklists, regulation or design methods, as these come with sections where risks of a model are addressed. For example,  the DEED could respond to the question \textit{Is  there  anything  about  the  composition of  the  dataset  or  the  way  it  was  collected  and preprocessed/cleaned/labeled that might impact future uses?} in datasheets for data sets \cite{gebruDatasheetsDatasets2020} or to the section on \textit{Caveats and recommendations} in model cards \cite{mitchell2019model}. It could be used to play out situations that would address AI regulation to see if data science teams have sufficient knowledge to apply these laws in their contexts. The drill would transform those hypothetical risks on paper into realistic problems which require a solution.

\subsection{Limitations}

There are some limitations to the studies we have carried out. As shown in the results, participants seemed to have good team spirit and were engaged in the topic. It is possible that the overwhelmingly positive feedback and the success of the discussions of the drill was due to participants being volunteers, who self-selected as being interested in the topic. We did not have participants who were resistant to change or to methods which explored data science topics in a non-technical manner. We cannot speak for teams with strong hierarchical structures either or in toxic workplaces, but we could foresee the DEED playing out in a radically different manner in such environments. Being open to challenging one's values and open to discussing value matters with colleagues is a core pillar of our work. While we believe it helps promoting ethical literacy in team discussions, we acknowledge that a DEED will not solve internal communication issues for some teams. 

Another non-trivial requirement from our participating teams is that they have the time and human resources to engage with the workshops. Resource limitations sometimes present a barrier to AI ethics uptake \cite{hopkinsMachineLearningPractices2021, varanasiItCurrentlyHodgepodge2023}. The suggested anatomy as we presented it in this paper would require at least half a day for the scenario crafting workshop and half a day for the drill execution and reflection. These workshops in themselves are not overly research intensive, but this does not take into account doing the work of implementing the findings of the drill. In our cases, the findings show that the drill discussions created enough momentum to implement additional responsible AI practices and to continue discussions in real life.

We recognize that some of the findings such as shifting focus or sharpening awareness for ethical issues are very "soft" in the sense that it is hard to verify if the participating teams were not developing these capabilities anyway or if they did so just by nature of engaging with a researcher, given the openness of the participating teams. While reflecting on process changes after the drill, Participant 1 said \textit{``...some of the learnings [of the DEED] are like very parallel to the [other projects], when we're kind of improving our processes. So it's just hard to decouple...''}. Since our current studies were all designed with participants being allowed to keep their participation anonymous, this may have hindered participants from using the DEED to directly advocate for any action point after the drill or talk to team members about their experience. In a less constrained setting, we would expect team members to communicate with each other directly after the drill (as opposed to a researcher presenting the findings) and potentially take ownership of any tasks that they identify with.

We have replicated the drill with three different sets of participants from two different commercial organisations, but the one constant in this drill has been the researchers designing prompt material for the scenarios and leading the workshops. Recent research has highlighted the important role of UX practitioners in navigating the complexities of responsible AI within tech \cite{wangDesigningResponsibleAI2023,zdanowskaStudyUXPractitioners2022}. It remains to be seen how easily technology professionals can lead the scenario crafting and drill sessions when they don't have the resources for external consultation. Most of the prompt artefacts presented in this drill, such as fake emails, are not difficult to produce and as the nature of the drill is to be close to reality, they do not require a particularly creative imagination either. 
Our intuition is that different designers, researchers, technologists or other AI/ML/DS stakeholders would only expand on the suggestions that we have provided. With the publication of this work, we hope to reach interested parties around the world and in different contexts to create Data Ethics Emergency Drills.

\balance
\subsection{Future Work}

One idea, that was outside the scope of these pilot studies, is to turn the drill into a conversation with the public outside of the AI industry. The DEED provides an obvious opportunity to not only engage with fictitious worries that exist inside a DS Team, but also to interact with concerns expressed by end-users, customers or wider society and to make them participants of the scenario crafting and the drill itself. Previous work on co-designing fictions with users and situated interventions with communities give an idea how this could be done successfully \cite{mullerExploringAIEthics2017, katell2020toward, rattaySensingCareDesign2023a, skirpanPrivacyCrisisExperienced2022}.

Furthermore, we would like to study more examples of teams completing a data ethics emergency drill, to gain a more general understanding of what types of data ethics dilemmas are addressed in scenarios and how this shapes their knowledge and understanding of responsible AI practice. These findings could add to existing literature of understanding ethical dilemmas in AI and ML \cite{dominguezhernandezToolkitDilemmasDebiasing2022}.

\section{Conclusion}

Developers have ``moved fast and broken things" for long enough. The Data Ethics Emergency Drill is an opportunity for them to take a step back from their application and reflect on their position and ability to react when things do break. It is a tool designed with and for data science practitioners that includes a workshop to brainstorm ethical emergencies that are relevant for their teams alongside proposed guidelines to execute emergency drills. 
Importantly, it is a proposal for new methods in responsible AI that elicit the linking of concrete work contexts to the abstract concepts of values and ethics. Participant feedback from initial runs of the DEED have shown that the format is realistic enough to foster useful discussions, help teams feel more comfortable in having challenging discussions, and elicit concrete action points to improve their processes with lasting impact in the teams conversations and responsible AI practice.

\begin{acks}
We would like to thank all the participants for their enthusiastic engagement with the DEED and invaluable input throughout the project. We would also like to thank Derek Edwards for the fantastic designs and Elaine Czech and Kenton O'Hara for their assistance with the paper. The first author was supported by the UKRI Centre for Doctoral Training in Interactive Artificial Intelligence (Grant Code: EP/S022937/1) with a studentship sponsored by LV= General Insurance. 
\end{acks}

\bibliographystyle{ACM-Reference-Format}
\bibliography{references}

\end{document}